\documentstyle[prd,aps,epsfig,floats]{revtex}
\begin{document}
\draft
\wideabs{
\title{BIASED METROPOLIS-HEAT-BATH ALGORITHM FOR FUNDAMENTAL-ADJOINT
       SU(2) LATTICE GAUGE THEORY}

\author{Alexei Bazavov$^{\rm \,a,b}$, Bernd A. Berg$^{\rm \,a,b}$,
        and Urs M. Heller$^{\rm c}$}

\address{ 
$^{\rm \,a)}$ Department of Physics, Florida State University,
  Tallahassee, FL 32306-4350, USA\\
$^{\rm \,b)}$ School of Computational Science, Florida State 
  University, Tallahassee, FL 32306-4120, USA\\
$^{\rm \,c)}$ American Physical Society, One Research Road,
  Box 9000, Ridge, NY 11961, USA\\ }

\date{\today }

\maketitle
\begin{abstract}
For SU(2) lattice gauge theory with the fundamental-adjoint action an
efficient heat-bath algorithm is not known so that one had to rely
on Metropolis simulations supplemented by overrelaxation. Implementing 
a novel biased Metropolis-heat-bath algorithm for this model, we find 
improvement factors in the range 1.45 to 2.06 over conventionally 
optimized Metropolis simulations. If one optimizes further with respect 
to additional overrelaxation sweeps, the improvement factors are found 
in the range 1.3 to 1.8.
\end{abstract}
}
\narrowtext

\section{Introduction}

Biased Metropolis Algorithms (BMAs) have been introduced quite some
time ago \cite{Ha70}, but they have not been applied beyond isolated 
classes of problems. Instead, the most frequently used Monte Carlo
schemes are the (original) Metropolis 
Algorithm (MA) and the Heat-Bath Algorithm (HBA), see \cite{BBook} for 
a textbook discussion. Both algorithms perform local updates of random 
variables, which, in lattice gauge theory, are matrices on the links of 
a 4D hypercubic lattice.

In its vanilla form, for lattice gauge theories, the MA proposes matrices
with the Haar measure of the gauge group. This suffers often from low
acceptance rates, but can be improved by restricting the proposal range
to a neighborhood of the matrix already in place.  However, one should
keep in mind that too small changes are not good either. A low acceptance
rate as well as too small changes lead to long autocorrelation times. As 
a general rule one should not tune up the acceptance rate to more than 
30\% to 50\% of the proposed updates (see, e.g., \cite{BBook}).

A way to improve the acceptance rate without restricting the proposals 
to a small range, and paying the price of large autocorrelation times,
is to perform multiple hits on the same matrix. As each hit, apart from
some common overhead, increases the CPU time needed linearly, an 
optimum is normally reached for a fairly small number of hits.

If one neglects CPU time requirements and counts only the number of 
link-updates the HBA achieves optimal performance in this class of 
local algorithms. By inverting the relevant cumulative distribution 
function it delivers the same results as a multi-hit Metropolis 
algorithm in the limit of an infinite number of hits per link update. 
This works very well in some cases, but in others the inversion is 
numerically so slow that, including CPU time in the balance sheet, 
a Metropolis scheme stays far more efficient than the HBA (which for 
many models has not even been constructed). 

In a previous paper \cite{BaBe05} two of the present authors have 
shown that the MA can  be biased so that it becomes an excellent 
approximation of the heat-bath updating, which was first introduced 
by Creutz \cite{Cr80} for SU(2) lattice gauge theory. The Biased 
Metropolis-Heat-bath Algorithm (BMHA) was illustrated for SU(2) 
and U(1) lattice gauge theories and the performance was found
competitive with the best implementations of the heat-bath algorithm
\cite{FH84,KP85,We89,HaNa92} for these models. In the present note 
we work out an example for which an efficient implementation of the 
conventional heat-bath algorithm does not exist: SU(2) lattice gauge
theory with the fundamental-adjoint action.

\section{The Model}

The SU(2) fundamental-adjoint action is 
\begin{equation} \label{e:act_fa}
  S(\{U\}) = \frac{\beta_f}{2} \sum_{\Box} {\rm Re}\,
  {\rm Tr} \left( U_{\Box}\right) + \frac{\beta_a}{3} \sum_{\Box}
  \left({\rm Re}\,{\rm Tr} \left( U_{\Box}\right)\right)^2\ .
\end{equation}
Here $U_{\Box} = U_{i_1j_1} U_{j_1i_2} U_{i_2j_2} U_{j_2i_1}$, where 
the sum is over all plaquettes of a 4D simple hypercubic $N_t\,N^3$
lattice, and $i_1,\,j_1,\,i_2,\,j_2$ label the sites circulating around 
the plaquette and $U_{ji}$ is the SU(2) matrix associated with the link 
$\langle ij\rangle$. The reversed link is associated with the inverse 
matrix. 

\begin{figure}[-t] \begin{center}
\epsfig{figure=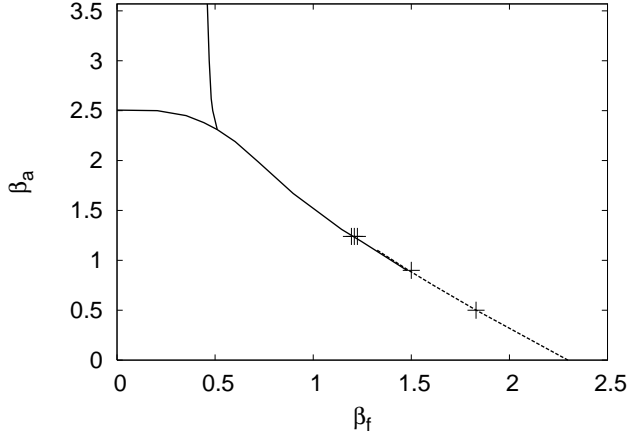,width=\columnwidth} \vspace{-1mm}
\caption{ Phase diagram of SU(2) lattice gauge theory with the
fundamental-adjoint action. The solid lines are the bulk transition 
and the dotted line indicates the $N_t=4$ deconfining transition. 
The five coupling constant values of our test runs are indicated by
$+$ signs. \label{fig_fa_betac} } \end{center} \vspace{-3mm} 
\end{figure}

This model has a bulk phase transition \cite{BhCr81} along lines in
the $(\beta_f,\beta_a)$ plane, see \cite{Ga97} and references given 
therein for more detailed investigations of this transition. 
Fig.~\ref{fig_fa_betac}, extracted from Ref.~\cite{BhCr81,Ga97}, 
shows the location of the bulk transition together with the $N_t=4$ 
deconfining phase transition line. While our aim is exclusively to
improve the MC algorithm, we target the phase transition lines,
interesting from the physics point of view and hence likely places
for future simulations, when choosing coupling constant values for
our test runs. Our test simulation points are also indicated in the
figure.

Our parameterization for SU(2) matrices is
\begin{equation} \label{SU2_parameters}
  U = a_0\,I + i\,\vec{a}\cdot \vec{\sigma},~~
  a_0^{\,2} + \vec{a}^{\,2} = 1,
\end{equation}
where $I$ denotes the $2\times 2$ identity matrix and $\vec{\sigma}$
are the Pauli matrices. A property of SU(2) group elements is that 
any sum of them is proportional to another SU(2) element. We define 
a SU(2) matrix $U_{\sqcup}$ which corresponds to a sum of the staples 
in equation (\ref{e:act_fa}) by
\begin{equation} \label{SU2_staple}
  s_{\sqcup}\,U_{\sqcup} = \sum_{k=1}^6 U_{\sqcup,k},~~
  s_{\sqcup} = \sqrt{\det\left(\sum_{k=1}^6 U_{\sqcup,k}\right)}\ .
\end{equation}
Here, $U_{\sqcup,k}$, $k=1,..,6$, denote the products of the three link
matrices which, together with $U$, the link to be updated, form one of
the six plaquettes containing the link to be updated.
The main step for implementing a BMHA is the table building process. 
Here we proceed in two steps. First, we use only the fundamental part 
and get the same table for the update variable $a_0$ and the parameter 
$s_{\sqcup}$ as in \cite{BaBe05}. This leads already to an increase
of the Acceptance Rate (AR) by a factor of nearly ten compared with 
the full-range MA. We refer to this approximation, which uses only 
the fundamental part of the action for table building, as BMHA-fund. 
As outlined in Ref.~\cite{BaBe05} such an updating table influences 
the efficiency through the AR, while the corresponding BMA is as 
exact as the usual MA. In Ref.~\cite{Hasenbusch} an essentially
equivalent algorithm was proposed and used for the fundamental-adjoint
SU(3) theory: do a Cabibbo-Marinari heatbath trial with the fundamental
part of the gauge action and the accept or reject with a Metropolis
step using the adjoint part of the action.

In a second step we construct our final BMHA by including the adjoint 
part of the action in a crude approximation, which is technically easy 
to handle and sufficient to increase the AR further. The amount of the
increase in AR is dependent on the two couplings (fundamental and
adjoint). At the critical endpoint of the bulk first order transition
line, for example, it is another 20\% over that
of the BMHA-fund to about 85\%.

To trace multiplicative factors clearly, we consider in the following
the action of our theory in $d$-dimensions. At each link we have 
$2(d-1)$ terms contributing to the $\Box$ sum. The link variables
are SU(2) matrices in the fundamental representation. A new link 
variable is proposed according to
\begin{equation}\label{e:Unew}
  U=U_rU_{\sqcup}^{-1}\,,
\end{equation}
where $U_r$ is randomly chosen with the appropriate measure and 
$U_{\sqcup}$ is a normalized staple matrix.

For constructing our BMHA we replace in the 
adjoint part each individual staple $U_{\sqcup,i}$ by
\begin{equation}\label{e:istaple}
  U_{\sqcup,i}\,\,\,\,\,\rightarrow\,\,\,\,\, \tilde U_{\sqcup,i}=
  \frac{1}{2(d-1)}\,U_{\sqcup}\,.
\end{equation}
This means, for the table we neglect individual staples fluctuations in 
the adjoint part. Instead of the adjoint part of the action we use
\begin{equation}\label{e:adj}
  \frac{\beta_a}{3} \sum_{i}^{2(d-1)} \left({\rm Re}\,{\rm Tr} 
  \left( Us_{\sqcup} \tilde U_{\sqcup,i}\right)\right)^2\,.
\end{equation}
Using (\ref{e:Unew}) and (\ref{e:istaple}) this reduces to
\begin{equation}\label{e:adj2}
  \frac{\beta_a}{3} \sum_{i}^{2(d-1)} \left({\rm Re}\,{\rm Tr} 
  \left( U_r\frac{s_{\sqcup}}{2(d-1)}\right)\right)^2\,.
\end{equation}
Nothing depends on the index of summation now, so the sum reduces to 
$2(d-1)$.  Also, as before ${\rm Re}\,{\rm Tr} (U_r)=2a_0$, and we get
for the adjoint contribution
\begin{equation}\label{e:adj3}
  \frac{\beta_a}{3}\,\frac{4a_0^2 s_{\sqcup}^2}{2(d-1)}.
\end{equation}
The total expression for the probability density which we tabulate is
\begin{equation} \label{Pa0}
  P(a_0) \sim \sqrt{1-a_0^{~2}}\,\exp\left(\beta_f\,s_{\sqcup}\,a_0+
  \frac{\beta_a}{3}\, \frac{4a_0^2 s_{\sqcup}^2}{2(d-1)} \right)\ .
\end{equation}
It has only one variable $a_0$ and one parameter $s_{\sqcup}$. As in 
\cite{BaBe05} we define $\alpha=\beta_f s_{\sqcup}$ for programming 
convenience. This substitution leads here to
\begin{equation}\label{Pa0fa}
  P(a_0) \sim \sqrt{1-a_0^{~2}}\,\exp\left(\alpha\,a_0+
  \frac{4}{3}\,\frac{\beta_a}{\beta_f^2}\,
  \frac{1}{2(d-1)}\,\alpha^2\,a_0^2 \right)\,.
\end{equation}

\section{Numerical Results}

To give an idea of how update proposals with a discretization 
of the probability density (\ref{Pa0fa}) work, we collect in 
table~\ref{t:small} the results of a short simulation on a $4^4$ 
lattice at $(\beta_f,\beta_a)=(1.5,0.9)$. The observables are 
the plaquette expectation values in the fundamental and adjoint 
representation, $\langle U_\Box^f\rangle$ and $\langle U_\Box^a
\rangle$, respectively, and their integrated autocorrelation times, 
$\tau_{int}(\langle U_\Box^f\rangle)$ and $\tau_{int}(\langle 
U_\Box^a\rangle)$. Here $\langle \cdot \rangle$ denotes the average
over a lattice configuration and $\langle\langle \cdot \rangle\rangle$ 
in table~\ref{t:small} the mean from all lattice configurations.
Autocorrelation times and error bars are calculated 
as explained in \cite{BBook}. While the estimated expectation values 
agree within statistical fluctuations, we find a dramatic increase 
of the AR from 6.5\% for the plain (full-range) MA to 62.4\% for 
the BMHA-fund and 85.2\% for the BMHA. This is accompanied by a 
decrease of the $\tau_{int}$ values, which is obvious for the 
first improvement step and within the limited statistics hardly 
visible for the second step (although certainly true due to the 
higher acceptance rate).

\begin{table} 
\caption{Simulation at $(\beta_f,\beta_a)=(1.5,0.9)$ on a $4^4$ 
         lattice relying on a statistics of 1000 sweeps for reaching 
         equilibrium and $32\times 1000$ sweeps with measurements. 
         Autocorrelation times are in units of MC sweeps.
          \label{t:small} }
\begin{tabular}{|l|c|c|c|}
          & Metropolis  & BMHA-fund & BMHA \\ \hline
$\langle\langle U_\Box^f\rangle\rangle$
          & 0.3451 (15) & 0.34636 (52) & 0.34694 (62)\\ \hline
$\langle\langle U_\Box^a\rangle\rangle$ 
          & 0.6368 (15) & 0.63798 (47) & 0.63853 (56)\\ \hline
$\tau_{int}(\langle U_\Box^f\rangle)$ 
          & 100.2 (8.6) & 19.5 (1.7)   & 19.8 (2.5)  \\ \hline
$\tau_{int}(\langle U_\Box^a\rangle)$
          &  95.9 (8.0) & 17.1 (1.4)   & 16.5 (2.2)  \\ \hline
AR in \%  &   6.5 (2)   & 62.4 (4)     & 85.2 (3)    \\
\end{tabular} \end{table}

\begin{table} \begin{center}
\caption{Relative efficiencies for our simulations on $4\times 8^3$ 
lattices. In columns 3--5 the efficiency of the BMHA over the
5-hit MA is shown for 0, 1 and 2 overrelaxation sweeps (plain, 1o 
and 2o). Columns 6 and 7 show how the BMHA is improved (or not) 
by additional overrelaxation hits. \label{t:efficiencies} }
\begin{tabular}{|c|r|r|r|r||r|r|}
 $(\beta_f,\beta_a)$ & AR & plain & 1o & 2o & 1obmha & 2obmha\\ \hline
 $(1.5,0.9)$    & 0.84 & 2.06 & 1.53 & 1.42 & 1.23 &  1.10  \\ \hline
 $(1.83,0.5)$   & 0.90 & 1.76 & 1.45 & 1.38 & 1.41 &  1.37  \\ \hline
 $(1.2146,1.25)$& 0.79 & 1.80 & 1.74 & 1.15 & 0.93 &  0.69  \\ \hline
 $(1.2,1.25)$   & 0.70 & 1.46 & 1.27 & 1.23 & 0.93 &  0.74  \\ \hline
 $(1.23,1.25)$  & 0.83 & 1.50 & 1.31 & 1.28 & 1.02 &  0.84  \\ \hline
\end{tabular} \end{center} \end{table}

Our main goal is to evaluate the BMHA against a MC algorithm, which 
was previously tuned by one of the authors for optimal 
performance~\cite{UH}. This is a $n$-hit Metropolis algorithm, with
update proposals by multiplying the old link matrix with an SU(2)
matrix centered around the unit element with a spread dynamically
adjusted to give an acceptance rate of about 50\% per Metropolis hit.
Doing 5 hits was found most cost effective.  For simulations on 
$4\times8^3$ lattices we compare the 5-hit MA with our BMHA 
implementation for several $(\beta_f,\beta_a)$ parameter values in the 
proximity of the bulk as well as the deconfining transition as shown 
in Fig.~\ref{fig_fa_betac} and compiled in table~\ref{t:efficiencies}.  
The AR of the BMHA is listed in column~2 of table~\ref{t:efficiencies}.  
Depending on the coupling constant values it varies in the range from 
70\% to 90\%. At all coupling constant values we have checked that the 
averages of our measured operators do, up to statistical fluctuation, 
not depend on the updating method.

\begin{table} \begin{center} 
\caption{Simulation at $(\beta_f,\beta_a)=(1.2146,1.25)$ on a 
$4\times 8^3$ lattice relying on a statistics of $2^{14}=16384$ 
sweeps for equilibration and $32\times 20480$ sweeps with measurements.
The CPU times are given in seconds. All other quantities are given in 
units of sweeps. \label{t:large} }
\begin{tabular}{|l|r|r|r|r|r|}
  & $\tau_{int}(\langle U_\Box^f\rangle)$ 
  & $\tau_{int}(\langle U_\Box^a\rangle)$ 
  & $\tau_{int}(\langle L\rangle)$ 
  & tunneling & $t_{CPU}$ \\ \hline
 5h   & 2294 (253) & 2262 (253) & 3430 (337) & 11.4 (1.2) $10^3$
                                             & 18390 \\ \hline
 1o5h & 1718 (137) & 1692 (136) & 2487 (238) & 5$\,$900 (420)
                                             & 25522 \\ \hline
 2o5h & 1209 (117) & 1193 (119) & 2131 (258) & 4$\,$667 (340)
                                             & 32292 \\ \hline
 bm   & 1804 (155) & 1776 (153) & 2981 (270) & 8$\,$323 (670)
                                             & 12958 \\ \hline
 1obm & 1222 (111) & 1204 (110) & 2083 (299) & 5$\,$190 (320)
                                             & 20573 \\ \hline
 2obm & 1172 (079) & 1156 (078) & 1746 (173) & 4$\,$520 (240)
                                             & 28295 \\
\end{tabular} \end{center} \end{table}

\begin{figure}[-t] \begin{center}
\epsfig{figure=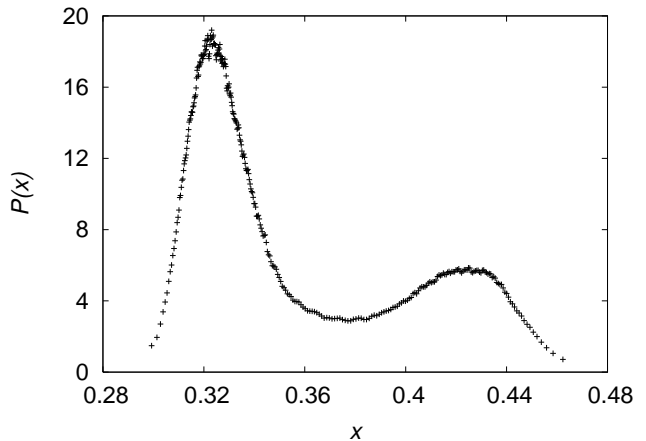,width=\columnwidth} \vspace{-1mm}
\caption{Probability density for the BMHA run (without overrelaxation)
of table~\ref{t:large}, $x=\langle U_\Box^f\rangle$. 
\label{fig_pdf_bw} } \end{center} \vspace{-3mm} \end{figure}

Including so-called overrelaxation sweeps \cite{Ad81,BrWo87,Cr87,HoKe98}
is known to reduce autocorrelation times, when the correlation length
becomes large, for example close to a second order phase transition.
For the fun\-da\-mental-adjoint
action, an exact overrelaxation step is not known to us. Instead we make
a trial overrelaxation update with the fundamental part of the action
and accept or reject the update according to the change in the adjoint
part of the action. The acceptance rate for these overrelaxation sweeps
decreases as the adjoint coupling becomes larger compared to the fundamental
coupling. For the couplings considered here the acceptance rate for the
overrelaxation sweeps varied between 69\% and 91\%.

In the subsequent tables, algorithms are encoded in the following way:
5h corresponds to the 5-hit Metropolis algorithm with the AR tuned to 
50\% per hit, bm to the BMHA, $i$o, with $i=1,2$ to doing 1 or 2
overrelaxation sweeps after each MA or BMHA update.

For $(\beta_f,\beta_a)=(1.2146,1.25)$ some of our data are compiled in
table~\ref{t:large}. This coupling constant point is pretty much on top
of the bulk $1^{st}$-order transition line, which leads to a double peak 
structure of the probability density of many observables, as illustrated 
in Fig.~\ref{fig_pdf_bw} for the plaquette expectation value in
the fundamental representation. Autocorrelation times of Polyakov 
loops $\langle L\rangle$ and tunneling times are also compiled in 
table~\ref{t:large}. Here the ``tunneling time'' is defined as the 
average number of sweeps the Markov process needs to propagate from 
one of the two maxima to the other and back. For all observables 
presented in table~\ref{t:large} we see that switching from the 5-hit 
MA to the BMA reduces not only the integrated autocorrelation and 
tunneling times, but also the CPU times.

The {\it efficiency} of an algorithmic approach~1 with respect to
an algorithmic approach~2 is given by
\begin{equation} \label{efficiency}
  E^{(1,2)} = \frac{\tau(2)_{int}}{\tau(1)_{int}}\,
           \frac{t(2)_{CPU}}{t(1)_{CPU}}\,.
\end{equation}
This formula reflects that the algorithm which needs less CPU time 
and produces a smaller value for the integrated autocorrelation time 
is the more efficient one. The $\tau(i)_{int}$ values, and hence the 
efficiencies, depend somewhat on the operator chosen. Using 
$\tau_{int}(\langle U_\Box^f\rangle)$ column~3 of 
table~\ref{t:efficiencies} collects the efficiencies found when 
comparing the BMHA with the 5-hit MA at our coupling constant values, 
as given in column~1.  Enhancements in the range 1.46 to 2.06
are found. Using other operators gives somewhat higher or lower 
efficiencies, but no systematic trend in either direction. For 
all operators we find always an improvement of the BMHA over the 
5-hit MA.

The values of column~3 of table~\ref{t:efficiencies} are reduced
by including overrelaxation sweeps in both the 5-hit MA and the BMHA
as is seen in columns~4 and~5.
This comes because the overrelaxation sweeps add uniformly CPU
time in both cases for which the integrated autocorrelation times 
get reduced by more or less the same fraction in case of the 5-hit
MA as well as for the BMHA. In all cases the numbers in columns~4 
and~5 of table~\ref{t:efficiencies} stay larger than 1, which means 
that the BMHA delivers always the better performance. 

The last point is to consider whether the increase of CPU time for 
including overrelaxation sweeps in the BMHA is justified by the 
achieved decrease of integrated autocorrelation times or not. This 
is done by calculating the efficiency of the BMHA with one or two 
overrelaxation sweeps with respect to the plain BMHA. The results 
are given in the last two columns of table~\ref{t:efficiencies}. 
We see that in two cases the performance with overrelaxation sweeps 
is worse (numbers $<1$) than for the plain BMHA. For another case 
there is almost no change, and in the two remaining cases one 
overrelaxation sweep (1o) before each BMHA sweeps is best. The 
points for which the overrelaxation sweeps help are close to the 
deconfining transition, where the correlation length is large and
overrelaxation sweeps are expected to be efficient,
whereas the other three points are close 
to the bulk transition.

In conclusion, while the need for overrelaxation sweeps varies, the 
BMHA outperforms the 5-hit MA always. Once constructed the BMHA 
does (in contrast to the 5-hit MA) not need any fine-tuning of 
parameters, so that it then provides a straightforward approach 
to performing pure lattice gauge theory simulations efficiently.

\acknowledgments
This work was in part supported by the US Department of Energy 
under contract DE-FG02-97ER41022. We thank Alexander Velytsky
for bringing, after submission of this paper, Ref.~\cite{Hasenbusch} 
to our attention.

\clearpage

\begin{thebibliography}{19}

\bibitem{Ha70} W.K. Hastings, Biometrika {\bf 57}, 97 (1970).

\bibitem{BBook} B.A. Berg, {\it Markov Chain Monte Carlo Simulations
and Their Statistical Analysis}, World Scientific, 2004.

\bibitem{BaBe05} A. Bazavov and B.A. Berg, Phys. Rev. D 
                 {\bf 71}, 114506 (2005).

\bibitem{Cr80} M. Creutz, 
Phys. Rev. D {\bf 21}, 2308 (1980).

\bibitem{FH84} K. Fabricius and O. Haan, 
Phys. Lett. B {\bf 143}, 459 (1984).

\bibitem{KP85} A.D. Kennedy and B.J. Pendleton, 
Phys. Lett. B {\bf 156}, 393 (1985).

\bibitem{We89} R.J. Wensley, Ph.D. Thesis, ILL-TH-89-25.

\bibitem{HaNa92} T. Hattori and H. Nakajima, Nucl. Phys. B (Proc. 
Suppl.) {\bf 26}, 635 (1992).

\bibitem{BhCr81} G. Bhanot and M. Creutz, 
                 Phys. Rev. D {\bf 24}, 3212 (1981).

\bibitem{Ga97} R.V. Gavai and M. Mathur, Phys. Rev. D {\bf 56}, 32 
        (1997).  

\bibitem{Hasenbusch} M. Hasenbusch and S. Necco, JHEP {\bf 0408},
        005 (2004). 

\bibitem{UH} U.M. Heller (unpublished). A similar SU(3) algorithm
was used for the work in
T. Blum, C. DeTar, U.M. Heller, Leo Karkkainen, K. Rummukainen and
D. Toussaint, Nucl.Phys. B {\bf 442}, 301 (1995); 
U.M. Heller, Phys.Lett. B {\bf 362}, 123 (1995). 

\bibitem{Ad81} S.L. Adler, Phys. Rev. D {\bf 23}, 2901 (1981).

\bibitem{BrWo87} F.R. Brown and T.J. Woch, Phys. Rev. Lett. {\bf 58},
2394 (1987).

\bibitem{Cr87} M. Creutz, Phys. Rev. D {\bf 36}, 515 (1987).

\bibitem{HoKe98} I. Horv\'ath and A.D. Kennedy, Nucl. Phys. B 
{\bf 510}, 367 (1998) and references therein. 

\end{thebibliography}
\end{document}